\documentstyle[12pt]{article}
\begin{document}

\def\gsim{{~\raise.15em\hbox{$>$}\kern-.85em
          \lower.35em\hbox{$\sim$}~}}
\def\lsim{{~\raise.15em\hbox{$<$}\kern-.85em
          \lower.35em\hbox{$\sim$}~}}
\def\epe{\varepsilon^\prime_{K}/\varepsilon_{K}}

\begin{titlepage}
\title{\Large{Supersymmetric Models with Approximate CP}}
\author{Galit Eyal\thanks{e-mail: galit@wicc.weizmann.ac.il}\\
\small{Department of Particle Physics, Weizmann Institute of Science,
Rehovot 76100, Israel}}
\date{\small{WIS-99/27/OCT-DPP}}
\maketitle

Supersymmetric models with an approximate CP, $10^{-3}\lsim \phi_{CP} \ll
1$, are a viable framework for the description of nature. The full high
energy theory has exact CP and horizontal symmetries that are
spontaneously broken with a naturally induced hierarchy of scales,
$\Lambda_{CP}\ll \Lambda_{H}$. Consequently, the effective low energy
theory, that is the supersymmetric Standard Model, has CP broken
explicitly but by a small parameter. The $\varepsilon_{K}$ parameter is
accounted for by supersymmetric contributions. The predictions for other
CP violating observables are very different from the Standard Model. In
particular, CP violating effects in neutral B decays into final CP
eigenstates such as $B\rightarrow \psi K_{S}$ and in $K\rightarrow
\pi\nu\bar{\nu}$ decays are very small. This framework, though, is
strongly disfavored by the recent measurements of $\epe$.
\end{titlepage}

\section{Introduction and Motivation}
Only two CP violating parameters have been measured to high accuracy so
far~\cite{a:pdg98}-\cite{a:na4899}:
\begin{eqnarray}
\varepsilon_{K} &=&(2.280\pm 0.013)\times 10^{-3} e^{i\frac{\pi}{4}}
\label{e:eKexp}\\
Re(\epe)&=&(2.11\pm 0.46)\times 10^{-3} \label{e:epeexp},
\end{eqnarray}
where
\begin{eqnarray}
\varepsilon_{K}&=&\frac{\left< (\pi\pi)_{I=0}|{\cal L}_{W}|K_{L}\right>}
{\left< (\pi\pi)_{I=0}|{\cal L}_{W}|K_{S}\right>},\\
Re(\epe)&=&\frac{1}{6} \left(\left|\frac{\left< \pi^{+}\pi^{-}|{\cal 
L}_{W}| K_{L}\right>}{\left< \pi^{+}\pi^{-} |{\cal L}_{W}|K_{S}\right>}
\frac{\left< \pi^{0}\pi^{0}|{\cal L}_{W}| K_{S}\right>}
{\left< \pi^{0}\pi^{0} |{\cal L}_{W}|K_{L}\right>}\right|^{2}-1\right).
\end{eqnarray}

Within the Standard Model (SM) the value of $\varepsilon_{K}$ can be
accounted for if the single CP violating phase, $\delta_{KM}$ in the
Cabibbo-Kobayashi-Maskawa (CKM) matrix, is of $O(1)$. Although the phase
is large, the effect is small due to flavor parameters. 

The theoretical interpretation of $\epe$ suffers from large hadronic
uncertainties. The SM theoretically preferred range is somewhat lower than
the experimental range (for recent work, see refs.~\cite{a:kns99,a:b99}
and references therein). Yet, if all the hadronic parameters are taking
values at the extreme of their reasonable ranges, the experimental result
can be accommodated.

There are CP violating observables that have not yet been accurately
measured, for example:
\begin{eqnarray}
a_{\psi k_{s}}\sin(\Delta m_{B}t)&=&
\frac{\Gamma(\bar{B}^{0}_{phys}(t)\rightarrow
\psi K_{S})-\Gamma(B^{0}_{phys}(t)\rightarrow \psi K_{S})}{\Gamma
(\bar{B}^{0}_{phys}(t)\rightarrow\psi K_{S})+\Gamma(B^{0}_{phys}(t)
\rightarrow \psi K_{S})}, \label{eq:apsiks}\\
a_{\pi\nu\bar{\nu}} &=&\frac{\Gamma(K_{L}\rightarrow \pi^{0}\nu\bar{\nu})}
{\Gamma(K^{+}\rightarrow \pi^{+}\nu\bar{\nu})}.
\end{eqnarray}
The values of these observables are predicted within the SM to 
be~\cite{a:babar,a:bb96}:
\begin{eqnarray}
(a_{\psi k_{s}})_{SM}&=& 0.4 -0.8,\\
(a_{\pi\nu\bar{\nu}})_{SM}&=& O(0.2).
\end{eqnarray}

The smallness of the measured parameters~(\ref{e:eKexp})-(\ref{e:epeexp}),
however, suggests that there might be new physics that allows a viable 
description of CP violating phenomena with approximate CP, that is with
all CP violating phases smaller than $O(1)$. In such a framework it is
possible that the predictions for other CP violating observables are
substantially different from the SM. In particular, $a_{\psi k_{s}}$ and
$a_{\pi\nu\bar{\nu}}$ are both much smaller than one.

Below we present a framework where the idea of approximate CP is realized.
This framework was introduced in ref.~\cite{a:en98}, where two explicit
supersymmetric (SUSY) models were given. Here, all the CP violating phases
are small. In particular $\delta_{KM}$ is small, and $\varepsilon_{K}$ is
accounted for by new physics, requiring at least one phase larger than or
of $O(10^{-3})$. We also report the results of a recent reexamination of
this framework~\cite{a:emns99}, in light of the accurate measurement of
$\epe$.

\section{The Framework}
Our high-energy theory is supersymmetric and has CP and abelian horizontal 
symmetries~\cite{a:fn79}. At low energies we assume that SUSY is softly
broken. Generic values for SUSY parameters might lead to too large flavor
changing neutral currents (FCNC). In our framework we use the breaking of
the horizontal symmetry supplemented by the mechanism of 
alignment~\cite{a:ns93} to avoid this problem. In order to account for CP
violation, we break CP spontaneously in such a way that in the low energy
effective theory the CP violating phases are small. With approximate CP, 
the potential CP problem of SUSY models, that is too large contributions
to electric dipole moments (EDM)~\cite{a:gnr97}, is avoided. 

Below we describe in more detail the various ingredients of our framework.

\subsection{Abelian Horizontal Symmetry}
Models of abelian horizontal symmetries are able to provide a natural
explanation for the hierarchy in the quark and lepton flavor
parameters~\cite{a:fn79,a:lns93}. The full high energy theory has an exact
horizontal symmetry, $H$. The superfields of the supersymmetric standard
model (SSM) carry $H$-charges. In addition there is usually at least one
SM singlet superfield, $S$, that also carries $H$-charge. The horizontal
symmetry is spontaneously broken when the SM singlet field assumes a
vacuum expectation value (vev), $\left< S\right>$. The breaking scale is
somewhat lower than a scale M where the information about this breaking is
communicated to the SSM, presumably by heavy quarks in vector like
representations of the SM (the Froggatt-Nielsen mechanism~\cite{a:fn79}).
The smallness of the ratio between the two scales, $\lambda \sim
\frac{\left< S\right>}{M}\ll 1$, is the source of smallness and hierarchy
in the Yukawa couplings. The parameter $\lambda$ is taken to be of order
of the Cabibbo angle, $O(0.2)$. Models are defined by the horizontal 
symmetry, the assigned horizontal charges and the hierarchy of vevs. For
most purposes it is sufficient to analyze the effective low energy theory,
which is the SSM supplemented with the following selection rules:

(i) Terms in the superpotential that carry charge $n$ under $H$ are
 suppressed by $\lambda^{n}$ if $n\geq 0$ and vanish otherwise.

(ii) Terms in the K\"ahler potential that carry charge $n$ under $H$ are
 suppressed by $\lambda^{|n|}$.

These selection rules allow estimation of the various entries in the
quark mass matrices $M^{q}$ and the squark mass-squared matrices
$M^{2}_{\tilde{q}}$ (the coefficients of $O(1)$ which appear in each entry   
are not known). The size of the bilinear $\mu$ and $B$ terms can also be
estimated. From the mass matrices, one can further estimate the mixing
parameters in the CKM matrix and in the gaugino couplings to quarks and
squarks. 

A convenient way to parameterize SUSY contributions to various processes is
by using the $(\delta_{MN}^{q})_{ij}$ parameters. In the basis where quark
masses and gluino couplings are diagonal, the dimensionless
$(\delta_{MN}^{q})_{ij}$ parameters stand for the ratio between
$(M^{2}_{\tilde q})^{MN}_{ij}$, the $(ij)$ entry ($i,j=1,2,3$) in the
squark mass-squared matrix ($M,N=L,R$ and $q=u,d$), and $\tilde m^2$, the
average squark mass-squared. If there is no mass degeneracy among squarks,
then these parameters can be related to the SUSY mixing angles. 

The naive values of the different parameters can be calculated using the
horizontal symmetry $U(1)$. For example, the naive estimate of the
$(\delta^{d}_{LR})_{12}$ parameter which is relevant to $\epe$ is given
by:
\begin{equation}
(\delta^{d}_{LR})_{12} \sim 
\frac{(M^{2}_{\tilde d})^{LR}_{12}}{\tilde{m}^{2}} \sim
\frac{\tilde{m}M^{d}_{12}}{\tilde{m}^{2}} \sim
\frac{m_{s}|V_{us}|}{\tilde{m}} \sim 
\lambda^{6}\frac{m_{t}}{\tilde{m}}.
\end{equation}

\subsection{Alignment}
The naive suppression of the supersymmetric flavor changing couplings is
not strong enough to solve all the SUSY FCNC problems. To solve the
$\Delta m_{K}$ problem, one can use the horizontal symmetry and holomorphy
to induce a very precise {\it alignment} of the quark mass matrices and
the squark mass-squared matrices~\cite{a:ns93,a:lns94}, resulting in a
very strong suppression of the relevant mixing angles in the gaugino
couplings to quarks and squarks. In order to achieve alignment, some of
the entries in $M^{d}$ should be suppressed compared to their naive
values. The required suppression is achieved by the use of holomorphy that
causes some of the Yukawa couplings to vanish~\cite{a:lns94}. In order to
achieve this, more than one $U(1)$ horizontal symmetry is required. These 
holomorphic zeroes are lifted when the kinetic terms are canonically 
normalized~\cite{a:lns94}, but their values are suppressed by at least a
factor of $\lambda^{2}$ relative to their naive value~\cite{a:en98}. 

Returning to our example we now find (horizontal symmetry + alignment):
\begin{equation}
(\delta^{d}_{LR})_{12} \sim \frac{\tilde{m}M^{d}_{12}}{\tilde{m}^{2}}
\lsim \lambda^{2}\frac{m_{s}|V_{us}|}{\tilde{m}} \sim
\lambda^{8}\frac{m_{t}}{\tilde{m}}.
\end{equation}

\subsection{Spontaneous CP Breaking}
As stated above, the high energy theory is CP symmetric. CP is
spontaneously broken in the following way. There are two SM singlet 
superfields, $S_{1}$ and $S_{2}$, that carry charges under the same $U(1)$ 
horizontal symmetry. Both of them receive vevs, $\left< S_{2}\right>\ll
\left< S_{1}\right>$. While one of the vevs can be chosen to be real, the
second is in general complex, with a phase of $O(1)$. The hierarchy
between the vevs and the relative, $O(1)$ phase, are naturally induced in
this framework~\cite{a:nr96}. This complex vev feeds down to all the 
couplings. 

In the low energy effective theory, there are many independent CP
violating phases, in particular in the mixing matrices of gaugino
couplings to fermions and sfermions. Furthermore, the ratio of vevs 
enables all CP violating phases to be suppressed, giving approximate CP.
The suppression of phases in the effective theory is by even powers of the
breaking parameter.

Returning to our example we find in this case (horizontal symmetry +
alignment + approximate CP):
\begin{equation}
Im(\delta^{d}_{LR})_{12} \lsim \lambda^{4}\frac{m_{s}|V_{us}|}{\tilde{m}}
\sim \lambda^{10}\frac{m_{t}}{\tilde{m}}. \label{e:delha}
\end{equation}

\section{Models and Predictions}
In ref.~\cite{a:en98} two representative models of approximate CP were
constructed. One of the models (model II) has the smallest viable CP
breaking parameter of $O(0.001)$, and the other (model I) has an 
intermediate value of $O(0.04)$. 

Regarding FCNC processes, we find in our models:

(i) The contributions to $\Delta m_{D}$ saturate the experimental upper   
 bound in both models. This is a generic feature of models of alignment,  
 related to the fact that in these models the Cabibbo mixing
 ($|V_{us}|\sim\lambda$) comes from the up sector.

(ii) The contributions to $\Delta m_{B}$ are very small.

(iii) The contributions to $\Delta m_{K}$ are of $O(10\%)$ in model I and 
 saturate the experimental value for model II. This is in contrast to all
 previous models of alignment where, to satisfy the $\varepsilon_{K}$
 constraint, SUSY contributions to $\Delta m_{K}$ were negligibly small.  
 
(iv) The contributions to other FCNC processes, such as $\Delta m_{B_{s}}$
 and $b\rightarrow s\gamma$, are very small. As concerns the rare
 $K^{+}\rightarrow\pi^{+}\nu\bar{\nu}$ decay, in both our models the SUSY
 contributions are of $O(10\%)$. While both the SM and the SUSY amplitudes
 are real to a good approximation, so that there is maximal interference
 between the two, the relative sign is unknown so that the rate could be  
 either enhanced or suppressed compared to the SM.

In both models $\varepsilon_{K}$ is accounted for by SUSY gluino-mediated
diagrams~\cite{a:ggms96}. Our results concerning CP violation are 
summarized in table~\ref{t:CPCP} where $a_{\psi K_{S}}$ and 
$a_{\pi\nu\bar{\nu}}$ are defined above, and $d_{N}$ is the EDM of the
neutron (given in units of $10^{-23} e\;cm$, so that the present
experimental bound is $d_{N}\lsim\lambda^{2}$).
\begin{table}[hbct]
\begin{center}
\begin{tabular}{|c|c|c|c|}
\hline
Process & SM & Model I & Model II \\
\hline
$a_{\psi k_{s}}$ & $O(1)$ & $O(\lambda^{2})\sim 0.04$ &
$O(\lambda^{4})\sim 10^{-3}$ \\
$a_{\pi\nu\bar{\nu}}$ & $O(\lambda)\sim 0.2$ & $O(\lambda^{4})\sim
10^{-3}$ & $O(\lambda^{8})\sim 10^{-6}$ \\
$d_{N}$ & 0 & $O(\lambda^{4})\sim 10^{-3}$ & $O(\lambda^{6})\sim 6\times
10^{-5}$ \\
\hline
\end{tabular} 
\end{center}
\caption{CP violating observables in the SM and in our models.}
\label{t:CPCP}
\end{table}

\section{$\epe$}
With approximate CP $\delta_{KM}$ is small and SM contributions can not 
account for the experimental measurement of $\epe$. New physics is
required. (If the relevant hadronic matrix element is much larger than its
value in the vacuum insertion approximation, as suggested by a recent
lattice calculation~\cite{a:bea99}, then the SM contribution with a small
value of $\delta_{KM}$ can account for $\epe$~\cite{a:em99}.) 

For SUSY to account for $\epe$, at least one of the following conditions
should be satisfied~\cite{a:ggms96},\cite{a:mm99}-\cite{a:s99}:
\begin{eqnarray}
Im[(\delta^d_{LL})_{12}]&\sim&\lambda\left(\frac{\tilde{m}}{500\;
GeV}\right)^2,\nonumber\\
Im[(\delta^d_{LR})_{12}]&\sim&\lambda^7\left(\frac{\tilde{m}}{500\;
GeV}\right),\\
Im[(\delta^d_{LR})_{21}]&\sim&\lambda^7\left(\frac{\tilde{m}}{500\;
GeV}\right),\nonumber\\
Im[(\delta^{u}_{LR})_{13}(\delta^{u}_{LR})_{23}^{*}]&\sim& \lambda^2, 
\nonumber\\
Im[V_{td}(\delta^{u}_{LR})_{23}^{*}]&\sim&\lambda^{3} 
\left(\frac{M_2}{m_W}\right),\\
Im[V_{ts}^*(\delta^u_{LR})_{13}]&\sim&\lambda^3\left(\frac{M_{2}}{m_{W}}
\right).\nonumber
\end{eqnarray}
In our framework only the conditions involving $(\delta^{d}_{LR})_{12}$
and $(\delta^{d}_{LR})_{21}$ can be met. Checking what are the lower
bounds on these parameters for extreme values of the parameters (for
details see ref.~\cite{a:emns99}) we find:
\begin{equation}
\label{eq:lowerLRot}
Im(\delta^{d}_{LR})_{12}\gsim 7\times 10^{-7},
\end{equation}
that is $O(\lambda^{9})$ or even $O(\lambda^{10})$ if $\lambda\sim
0.24$. A similar bound applies to $Im[(\delta^{d}_{LR})_{21}]$.

In models in which the flavor problems are solved by alignment, but the CP
problems are solved by approximate CP, eq.~(\ref{e:delha}) holds. This is
consistent with the experimental constraint of eq.~(\ref{eq:lowerLRot}) 
only if all the following conditions are simultaneously satisfied:

(i) The suppression of the relevant CP violating phases is `minimal',
 that is $O(\lambda^{2})$.

(ii) The alignment of the first two down squark generations is `minimal', 
 that is $O(\lambda^{2})$.

(iii) The mass scale of the SUSY particles is low, $\tilde{m} \sim 150\; 
 GeV$.

(iv) The hadronic matrix element is larger than what hadronic models
 suggest.

(v) The mass of the strange quark is at the lower side of the
 theoretically preferred range.
 
(vi) The value of $\epe$ is at the lower side of the experimentally
 allowed range.\\
We conclude that models that combine alignment and approximate CP are
disfavored by the measurement of $\epe$. More than that, the explicit
models (model I and II) described above are ruled out by this
measurement.

We do note, however, that models of abelian horizontal symmetries and
approximate CP where the flavor problems are solved by a mechanism
different from alignment can account for $\epe$.

\section{Conclusions}
In the near future, we expect first measurements of various CP asymmetries 
in $B$ decays, such as $B\rightarrow\psi K_S$ or $B^\pm\rightarrow\pi^0
K^\pm$. If these asymmetries are measured to be of order one, it will
support the SM picture, that the CP violation that has been measured in 
the neutral $K$ decays is small because it is screened by small mixing
angles, while the idea that CP violation is small because all CP violating  
phases are small will be excluded. It is interesting, however, that 
various specific models that realize the latter idea, such as those
discussed in this work, can already be excluded by the measurement of a 
tiny CP violating effect, $\varepsilon'_{K}\sim 5\times10^{-6}$.\\

{\bf Acknowledgments}\\
I thank Antonio Masiero, Yossi Nir and Luca Silvestrini for enjoyable
collaborations on the topics presented here.

{}

\end{document}